# GLUON LOOPS IN THE INELASTIC PROCESSES IN QCD

I. V. Sharf[1], K.K. Merkotan[1], N.A. Podolyan[1], D.A. Ptashynskyy[1], A.V. Tykhonov[2], M.A. Deliyergiyev[2], G.O. Sokhrannyi[1], V.D. Rusov[1,3]

[1] *Department of Theoretical and Experimental Nuclear Physics, Odessa National Polytechnic University*

[2] *Department of Experimental Particle Physics, Jozef Stefan Institute, Jamova 39, SI-1000 Ljubljana, Slovenia*

[3] *Department of Mathematics, Bielefeld University, Universitatsstrasse 25, 33615 Bielefeld, Germany*

Abstract: It is shown that inelastic process of the exchange with two massless gluons is formally equivalent to the process of the exchange with one massive particle. Thus, using the Laplace's method, a new mechanism of mass generation in inelastic processes is discovered, which is described by the non-Abelian gauge theory. Furthermore, it is shown that in the QCD perturbation theory, the same mechanisms of cross-sections growth take place, similar to the ones discovered before in the effective scalar theories.

## 1. Introduction

In previous works [1-5] the permissibility of applying the Laplace's method to calculation of the inelastic scattering cross-sections has been demonstrated on the example of Feynman diagrams for effective scalar field theories. The aim of the work is the application of this method to a more realistic theory, i.e. QCD. At this point the problem occurs due to the fact that within QCD perturbation theory one considers the diagrams of interacting quarks and gluons, while the particles in the initial and final states represent the bound states of quarks and gluons. Usually this problem is overcome by representing the state of interacting hadrons via parton distribution functions, which is the result of bringing into the theory the large number of adjustable parameters.

However, it is well known that the static hadrons can be described within the non-relativistic quark models. That is, in rest frames of hadrons in the initial and final states they can be considered in the non-relativistic approximation. In this approximation, hadrons can be treated as those consisting of a given number of quarks. Moreover, their interaction can be described by means of potential energy, ignoring the existence of gluon fields in hadrons.

The usual formulation of scattering problem is that the particles in the initial and final states do not interact with each other. However, for the quarks inside the initial and final hadrons we cannot neglect the interaction between them, because it keeps them bound. In the perturbation theory we consider the problem in interaction representation. Non-relativistic interaction operator in the interaction representation is described as an integral of oscillating function, which according to the Riemann - Lebesgue theorem, should be close to zero when the time approaches the plus and minus infinity, that corresponds to the initial and final states. Therefore, we can suppose the initial and final quarks to be free, though they are bound.

In the non-relativistic approximation hadron's mass must be equal to the sum of the masses of component quarks. If we neglect all interactions comparing to the strong one, these masses can be considered approximately equal between themselves, i.e. we assume that the hadron mass is divided into equal parts between the masses of quarks. So, for example, for baryons we consider quarks with the "effective" mass equal to one third of the hadron mass in the rest frame.

It is found that in the hadron rest frame its energy momentum four-vector is divided equally between quarks, but then turning to an arbitrary frame, we get that in the initial and final states hadrons consist of effectively free quarks, each of which carries either third part (for baryons) or half part (for mesons) of the hadron energy momentum four-vector.

So, we are going to consider a model in which the initial and final state consists of effectively free quarks, each of which carries either the third part or the half part of the momentum of "its own" hadron.

The simplest diagram of the charged mesons formation in this model is the diagram shown in Fig.1a.

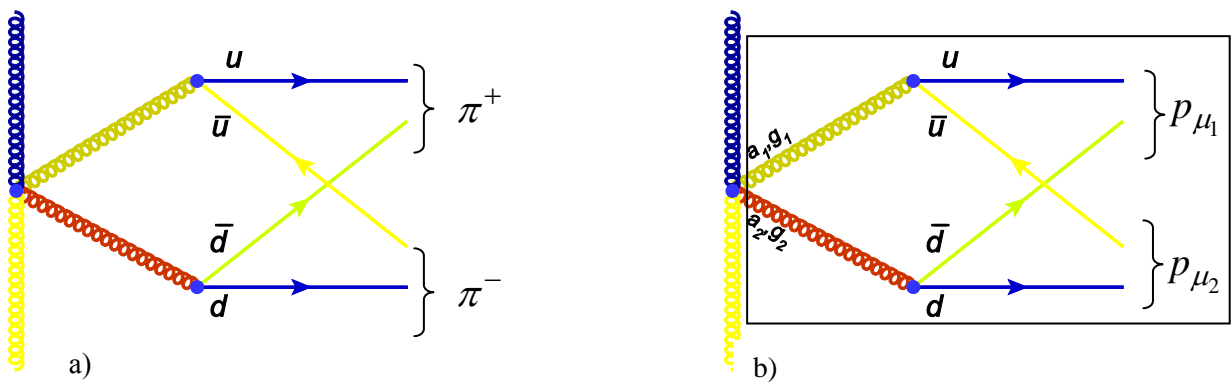

Fig.1. a) The simplest diagram of the charged mesons formation. b) The part of the diagram that corresponds to the hadrons formation ($a_1, a_2$ - Lorentz indices of gluons, $g_1, g_2$ - internal indices).

Thus, if we separate a part of the diagram as in the fig.1b, then the expression, that transforms as two-index tensor under the Lorentz transformations (with indices $a_1$ and $a_2$) and also as two-index tensor under transformations of adjoint $SU(3)$ group representation (with indices $g_1$ and $g_2$), corresponds to this part. The tensor should be contracted with Lorentzian and internal tensor corresponding to the four-gluon vertex in fig.1a,b. The expression for the scattering amplitude includes the contraction as a multiplier, which is a scalar function $f\left(p_{\mu_1}, p_{\mu_2}\right)$ of these groups changes the function of mesons four-momenta $p_{\mu_1}$ and $p_{\mu_2}$ in fig.1b under both these transformation groups.

For the Laplace method application we need to find the maximum point of the scattering amplitude modulus squared. It should be $p_{\mu_1} = p_{\mu_2} = p$ at the maximum point because of the finite mesons diagram symmetry. Therefore, searching for the maximum we can consider the contraction $f_1(p) = f\left(p_{\mu_1} = p, p_{\mu_2} = p\right)$. Taking into account a scalar value of the function $f_1(p)$, we draw a conclusion that it should be a function of the scalar square $p^2$, because it is the only scalar combination that can be formed from the four-vector components. Therefore, due to the mass shell condition for particles in the final state, this function reduces to a constant. Thus the scattering amplitude maximum modulus achieves at such a contraction that a constant can match the part of the diagram in fig. 1a.

We note that the above reasoning can be applied for any type of diagram shown in fig.2a. Therefore, at the stage of maximum searching it can be replaced by a constant that does not influence either the presence or the position of the maximum. So, to simplify further image in the diagrams, we will show the unit as in fig.2b.

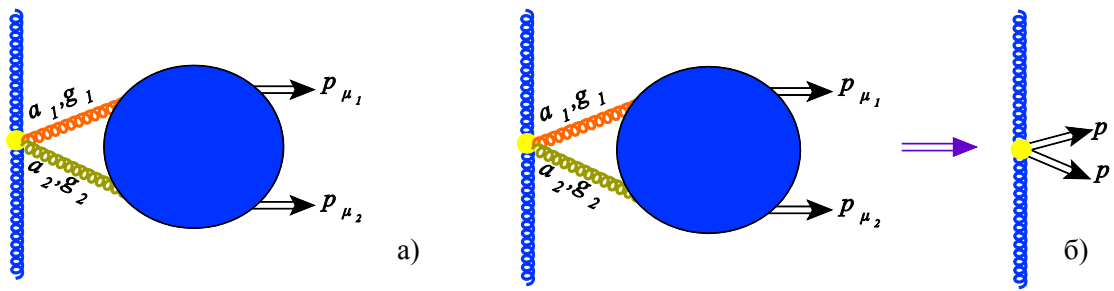

Fig.2. a) General diagram of the hadrons formation, b) Its replacement by the effective vertex.

Also in QCD due to the color conservation law it is impossible to consider inelastic scattering diagrams with one gluon exchange (fig.3a). Formally, it is revealed that calculating a color indices sum corresponding to the diagram contribution to the scattering amplitude, we get zero.

The simplest version of the diagram, for which the color indices sum gives a nonzero result, is shown in fig.3b. But it is necessary to consider such diagrams because that the proton-proton scattering process cannot be described by a model with tree diagrams (i.e. those that do not contain loops), and we should consider diagrams with gluon loops.

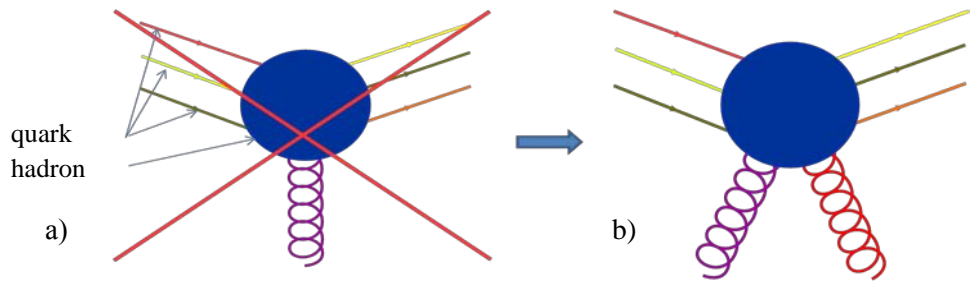

Fig.3. a) The forbidden process, in which the colorless system of three quark exchanges one gluon with another system. b) A simple diagram "allowed" by color conservation.

The simplest diagrams of the inelastic scattering with gluon loops for solution of these problems are shown in fig.4.

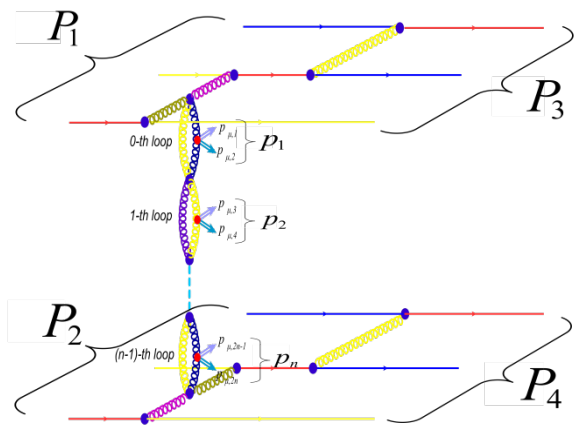

Fig. 4. The simplest diagram of the inelastic scattering with gluon loops.

The further simplification is that the diagrams in fig.4 correspond to the analytical expressions, which are the product of multipliers that are matched to each of two proton units (top and bottom in fig.4) and also a multiplier, that is matched a set of gluon loops (fig.5).

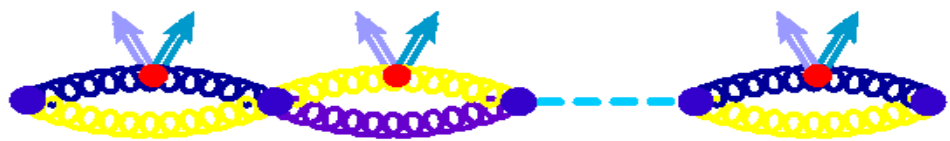

Fig.5. Inside part of diagram in fig.4, which contains a set of gluon loops.

It allows exploring and maximizing modulus of these multipliers separately. In addition, an expression that is a product of multipliers, each of which corresponds to the single gluon loop, corresponds to the part of the diagram in fig.5. The integration by four-momenta that circulate through loops can also be performed separately from each other.

The aim of this work is to find the maximum point of the scattering amplitude modulus squared and to study its properties for the inside parts of the diagram (fig.4), shown in fig.5.

### 1. Calculation of the gluon loop integral.

According to Feynman's diagram technique, up to a constant term, the expression corresponding to the gluon loop (Fig. 6) has the form:

$$A = \int d^4q \frac{1}{(K-q)^2 + i\varepsilon} \frac{1}{q^2 + i\varepsilon} \frac{1}{(q-p)^2 + i\varepsilon} \qquad (1)$$

Hereinafter the sum is denoted by $p$

$$p = p_{\mu_1} + p_{\mu_2} \qquad (2)$$

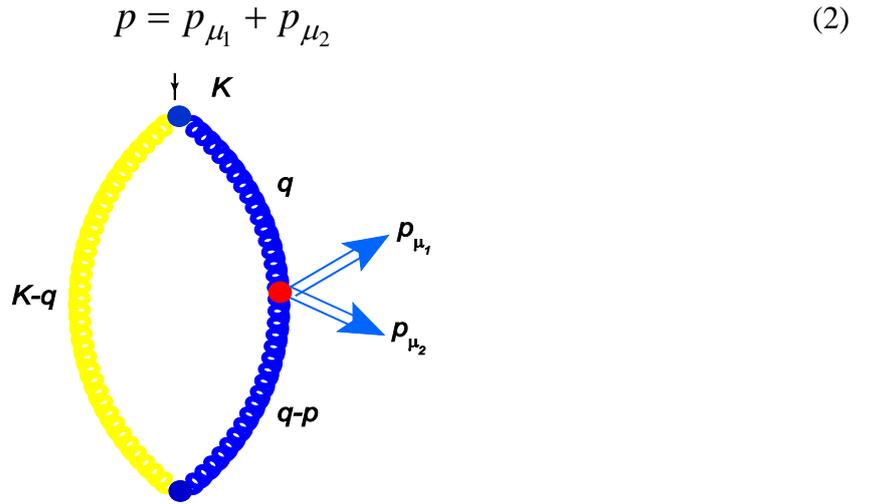

Fig.6 Four-momentum lines of a gluon loop

Similarly to what has been proved in Ref. (1), we can show that the scalar squares of four-momentum $K$, entering each gluon loop, are negative.

After applying the Feynman identity to Eq.(1), and performing the transformation, it can be reduced to a two-dimensional integral over the triangle $\triangle ABC$ of the Fig.7:

$$A = \frac{\pi^2 i}{p^2 \sqrt{3}} \iint_{\triangle ABC} dy_1 dy_2 \frac{1}{\left(-\sqrt{\frac{2}{3}} y_1 + \frac{Kp}{p^2}\left(\sqrt{\frac{1}{6}} y_1 - \frac{1}{\sqrt{2}} y_2 + \frac{1}{3}\right) - \frac{1}{6}\right)^2 - b\left(\sqrt{\frac{1}{6}} y_1 - \frac{1}{\sqrt{2}} y_2 + \frac{1}{3} - l\right)^2 + d + i\varepsilon}, \qquad (3)$$

where

$$b = \frac{(Kp)^2 - K^2 p^2}{(p^2)^2}, d = \frac{1}{4}\left(\frac{(Kp - K^2)^2}{(Kp)^2 - K^2 p^2} - 1\right), l = \frac{p^2(Kp - K^2)}{2((Kp)^2 - K^2 p^2)}. \quad (4)$$

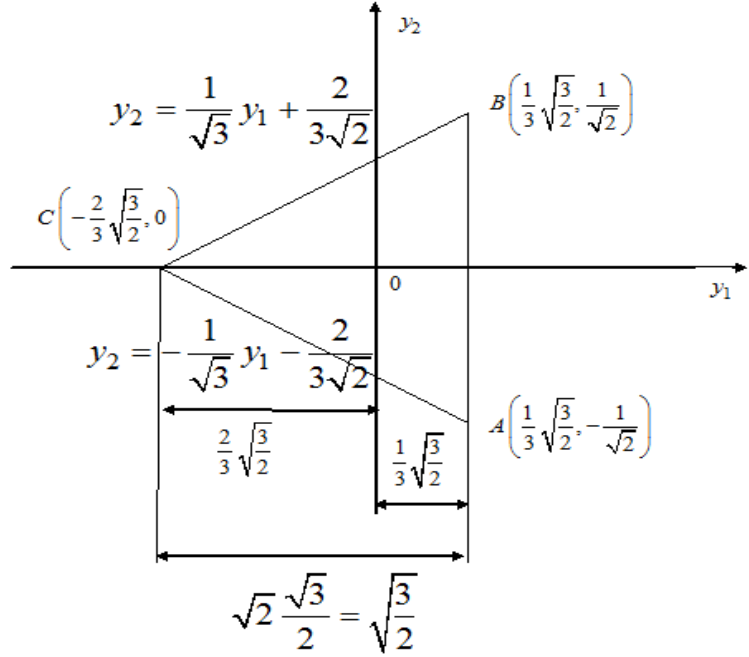

Fig.7 The region of integration in (3).

As one can see from Eq. (4), considering the negativity of scalar square of four-vector $K$, entering the loop (fig.6), we have $b > 0$. That means the quadratic form in the denominator of the integral (3) is the difference of squares. Therefore, for the further calculation of the integral (3), it is convenient to pass to new variables

$$u = \left(-\sqrt{\frac{2}{3}} y_1 + \frac{Kp}{p^2}\left(\sqrt{\frac{1}{6}} y_1 - \frac{1}{\sqrt{2}} y_2 + \frac{1}{3}\right) - \frac{1}{6}\right) - \sqrt{b}\left(\sqrt{\frac{1}{6}} y_1 - \frac{1}{\sqrt{2}} y_2 + \frac{1}{3} - l\right),$$

$$v = \left(-\sqrt{\frac{2}{3}} y_1 + \frac{Kp}{p^2}\left(\sqrt{\frac{1}{6}} y_1 - \frac{1}{\sqrt{2}} y_2 + \frac{1}{3}\right) - \frac{1}{6}\right) + \sqrt{b}\left(\sqrt{\frac{1}{6}} y_1 - \frac{1}{\sqrt{2}} y_2 + \frac{1}{3} - l\right). \quad (5)$$

In these variables, instead of expression (3) we get:

$$A = \frac{-\pi^2 i}{2\sqrt{(Kp)^2 - K^2 p^2}} \iint_{\triangle ABC} dudv \frac{1}{uv + c^2 + i\varepsilon}, \quad (6)$$

where the following notation is used:

$$c^2 = \frac{K^2(K-p)^2}{4((Kp)^2 - p^2 K^2)} > 0. \tag{7}$$

The triangle $\square ABC$ maps on the triangle which is shown in Fig.8:

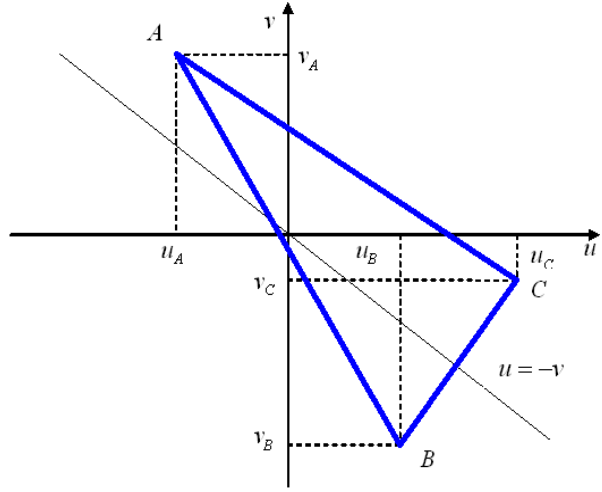

Fig.8 The region of integration in (6).

Coordinates of vertices C and B of the triangle fig.7 expressed through the outer four-momentum of loop in fig.6:

$$u_C = \frac{1}{2} + \frac{1}{2}\frac{Kp - K^2}{\left(\sqrt{(Kp)^2 - K^2 p^2}\right)} > 0, v_C = \frac{1}{2}\left(1 - \frac{Kp - K^2}{\sqrt{(Kp - K^2)^2 - K^2(K-p)^2}}\right) < 0,$$

$$u_B = -\frac{1}{2} + \frac{1}{2}\frac{(Kp - K^2)}{\sqrt{(Kp)^2 - K^2 p^2}} = -v_C > 0, v_B = -\frac{1}{2} - \frac{1}{2}\frac{(Kp - K^2)}{\sqrt{(Kp)^2 - K^2 p^2}} < 0. \tag{8}$$

The expression for the coordinate $v_A$ can be written in the form:

$$v_A = \frac{1}{2}\left(\frac{K^2 - (K-p)^2}{p^2} + \frac{\left(\frac{K^2 - (K-p)^2}{p^2}\right)^2 - \left(\frac{K^2 + (K-p)^2}{p^2}\right)}{\sqrt{1 - 2\left(\frac{K^2 + (K-p)^2}{p^2}\right) + \left(\frac{K^2 - (K-p)^2}{p^2}\right)^2}}\right). \tag{9}$$

Taking into account the negativity of $K^2$ and $(K-p)^2$, one can see that second term in (9) is positive. After the calculation of the difference of two squares terms in parentheses, we get:

$$\left(\frac{\left(\dfrac{K^2-(K-p)^2}{p^2}\right)^2-\left(\dfrac{K^2+(K-p)^2}{p^2}\right)}{\sqrt{1-2\left(\dfrac{K^2+(K-p)^2}{p^2}\right)+\left(\dfrac{K^2-(K-p)^2}{p^2}\right)^2}}\right)^2-\left(\frac{K^2-(K-p)^2}{p^2}\right)^2=\frac{4K^2(K-p)^2}{(p^2)^2}>0. \quad (10)$$

For this we draw a conclusion, that the absolute value of second positive term in parentheses (9) is greater than the first one. Therefore, in spite of the sign of the first term, we have $v_A > 0$.

The expression for the coordinate $u_A$ has the form:

$$u_A = \frac{1}{2}\left(\frac{K^2-(K-p)^2}{p^2}-\frac{\left(\dfrac{K^2-(K-p)^2}{p^2}\right)^2-\left(\dfrac{K^2+(K-p)^2}{p^2}\right)}{\sqrt{1-2\left(\dfrac{K^2+(K-p)^2}{p^2}\right)+\left(\dfrac{K^2-(K-p)^2}{p^2}\right)^2}}\right), \quad (11)$$

which differs from the Expression (9) for $v_A$ only by its sign. Now, the largest modulus in the parentheses of (11) is negative. Therefore $u_A < 0$.

As will be shown below, the Jacobian of the transformation (5) plays a significant role for each of the loops Fig.5. The expression for the Jacobian of this transformation has the form

$$J = \frac{\partial(u,v)}{\partial(y_1,y_2)} = \frac{2}{\sqrt{3}}\frac{\sqrt{(Kp)^2-K^2p^2}}{p^2}. \quad (12)$$

Let's consider the further calculation of the integral (6). The real part of the denominator of the integrand expression in (6) turns to zero along the hyperbola

$$uv + c^2 = 0. \quad (13)$$

Inserting Eqs. (8), (9) and (11) into this expression demonstrates that the hyperbola includes all three vertices of the triangle $\square ABC$ in Fig. 8, see Fig. 9. As a result, within the region, bounded by the segments $AB$ and $AC$ and the arc of hyperbola $BC$, one gets the inequality

$$uv + c^2 > 0, \quad (14)$$

and in the region between the arc $BC$ and segment $BC$:

$$uv + c^2 < 0. \quad (15)$$

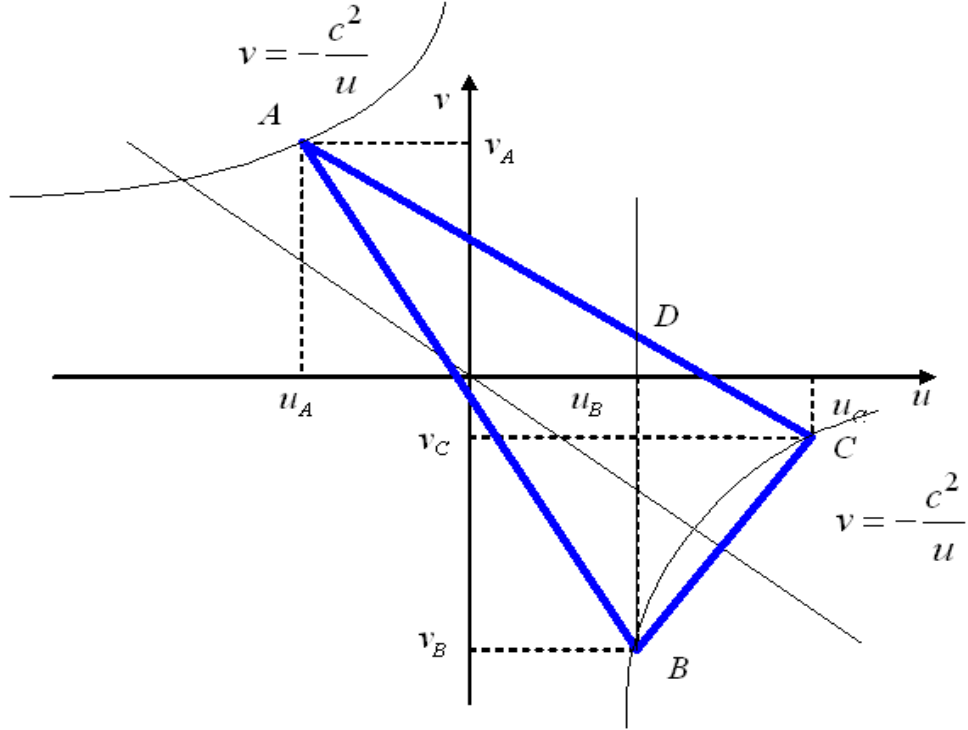

Fig.9 The relative position of the integration domain in Eq.(6) and hyperbola Eq.(13), on which the real part of the denominator in (6) changes its sign.

Hence, due to the presence of region, in which the relations (15) take place, multiplier $A$, which corresponds to each loop in Fig.5, and which is expressed by Eq. (6), has a nonzero imaginary part. It can be represented as follows:

$$A = \frac{i\pi^2}{2} A',$$

$$\text{Re}(A') = \frac{1}{\sqrt{(Kp)^2 - K^2 p^2}} \left( \int_{u_A}^{u_B} \frac{1}{u} \ln\left( \frac{k_{AC}(u - u_C)}{k_{AB}(u - u_B)} \right) du + \int_{u_B}^{u_C} du \frac{1}{u} \ln\left( \frac{k_{AC}(u_A - u)}{(u - u_B)} \right) \right), \quad (16)$$

$$\text{Im}(A') = \frac{\pi}{\sqrt{(Kp)^2 - K^2 p^2}} \ln\left( \frac{u_C}{u_B} \right).$$

Here $k_{AB}$ and $k_{AC}$ denote the angular coefficients of the lines $AB$ and $AC$ in Fig.9.

The expression (16) includes one-dimensional integrals, which can be calculated numerically. This allows to numerically calculate the multipliers, each of which corresponds to one of the loops in Fig. 5, and to numerically maximize the square modulus of their product.

## 2. The description of the maximization procedure.

Consider scattering amplitude that corresponds to the diagram Fig. 5 in c.m.s. of secondary particles. Similar to [1-4] one can demonstrate that the search for the maximum of the scattering amplitude squared modulus can be reduced to such a domain of momenta space, where the transverse components of secondary particles four-momenta are zero. The conservation law of transverse momenta is automatically fulfilled in this domain. Therefore, one has to fulfill only the energy and longitudinal momentum conservation law. The longitudinal components of the particles momenta $P_1$ and $P_3$ can be found from these conservation laws (Fig. 4). It should be noted that, like in [2,5], all expressions are nondimensionalized with the mass of secondary particle, in our case the mass of pion. Moreover, instead of longitudinal components of the secondary particles momenta, we use rapidities as independent variables of the scattering amplitude (Fig. 4):

$$p_{a,\square} = 2\operatorname{sh}(y_a), a = 1, 2, \ldots, n. \quad (17)$$

Using all these notations, from the energy-momentum conservation law it follows that

$$P_{3\square} = \frac{1}{2}\left(-P_{\square p} + E_p\sqrt{1 - \frac{4M^2}{\left(E_p\right)^2 - \left(P_{\square p}\right)^2}}\right), \quad P_{4\square} = \frac{1}{2}\left(-P_{\square p} - E_p\sqrt{1 - \frac{4M^2}{\left(E_p\right)^2 - \left(P_{\square p}\right)^2}}\right), \quad (18)$$

where:

$$\sum_{a=1}^{n} 2\operatorname{sh}(y_a) = P_p, \quad \sqrt{s} - \sum_{a=1}^{n} 2\operatorname{ch}(y_a) = E_p, \quad (19)$$

here $M$ is dimensionless proton mass.

Substituting (18) into the expression for scattering amplitude and making use of the fact, that in considered domain of phase space all transverse momenta are zero, we can calculate all four-vectors $K$, entering each loop of Fig. 4, and consequently their Lorenz-invariants $K^2$ and $(Kp)$ as functions of rapidities (17).

Next, using (8)-(16) we can subsequently calculate the multipliers corresponding to the loops in Fig. 4 as functions of rapidities, and therefore calculate the amplitude, which correspond to Fig. 5. Let's denote it through $A(y_1, y_2, \ldots, y_n)$. After that we can numerically maximize the logarithm of this expression.

In order to control the maximization of the scattering amplitude we use functions $G_a^n(x)$, which we define in the following way. Let's denote the particle's rapidities that maximize the scattering amplitude through $y_1^{(0)}, y_2^{(0)}, \cdots, y_n^{(0)}$, then

$$G_a^n(x) = \left|A\left(y_1^{(0)}, y_2^{(0)}, \cdots, y_{a-1}, y_a + x, y_{a+1}, \cdots y_n^{(0)}\right)\right|^2, a = 1, 2, \ldots, n. \tag{20}$$

It can be shown that each function $G_a^n(x)$ satisfies the following expressions:

$$G_a^n(\Delta x) < G_a^n(0), G_a^n(-\Delta x) < G_a^n(0), a = 1, 2, \ldots, n. \tag{21}$$

for any insignificant value of $\Delta x$. The validity of these expressions for any $a$ means, that all first partial derivatives of function $|A(y_1, y_2, \ldots, y_n)|^2$ in the point $\left(y_1^{(0)}, y_2^{(0)}, \cdots, y_{a-1}^{(0)}, y_a^{(0)}, y_{a+1}^{(0)}, \cdots y_n^{(0)}\right)$ are zero. The plots for some of these functions $G_a^{n=15}(x)$, maximized for energy $\sqrt{s} = 15 GeV$, are presented in the Fig. 10.

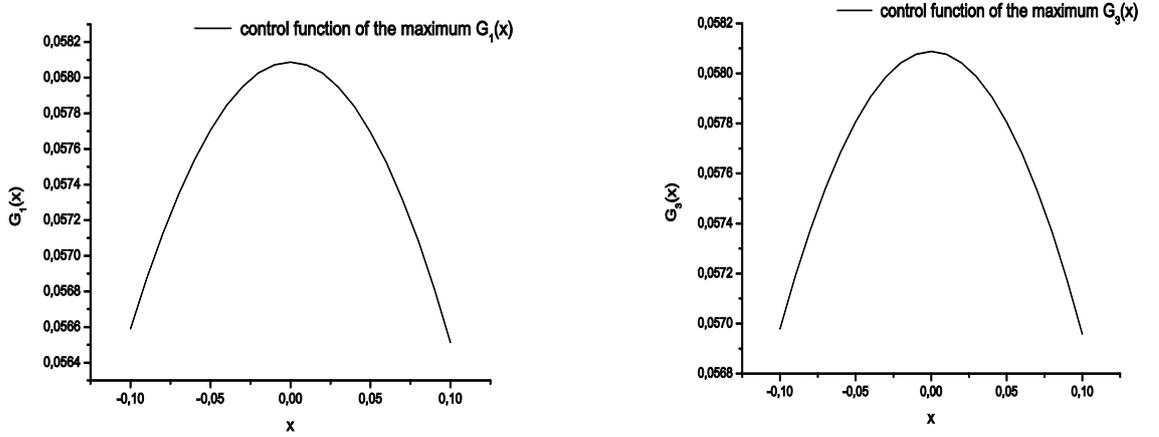

Fig.10. Typical plots of the function (20) that has been used to control the maximization.

The following functions were used for 3D visualization of maximum point:

$$G_{ab}^n(u,v) = \left|A\left(y_1^{(0)}, \ldots, y_{a-1}^{(0)}, y_a^{(0)} + u, y_{a+1}^{(0)}, \ldots, y_{b-1}^{(0)}, y_b^{(0)} + v, y_{b+1}^{(0)} \ldots, y_n^{(0)}\right)\right|^2, \tag{22}$$
$$a = 1, 2, \ldots, n, b = 1, 2, \ldots, n.$$

Some of them are presented here in Fig. 11.

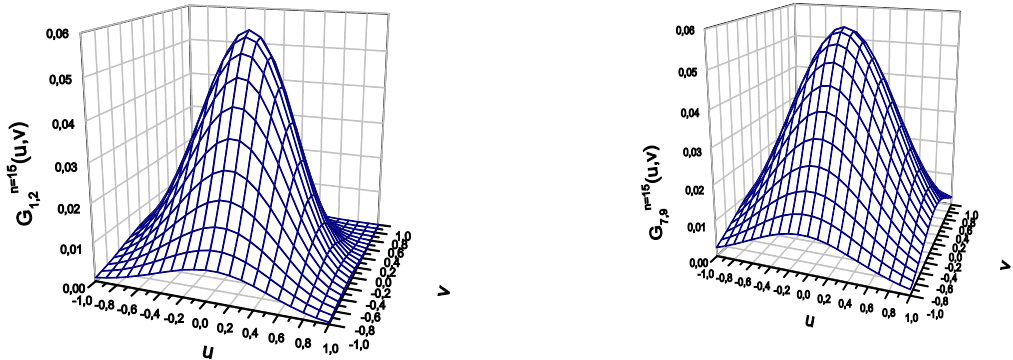

Fig. 11 Typical plots of the function (22)

Hence, the scattering amplitude that corresponds Fig. 5, has the constrained maximum when the energy-momentum is conserved. Let us consider now some important properties of this maximum.

### 3. The results of numerical maximization.

The results of the numerical maximization of the squared modulus of scattering amplitude Fig. 5 appeared to be very similar to the results for a diagram in simple scalar models, obtained in [1,5]. Moreover, as one can see from Fig. 12, the values of the scattering amplitude squared modulus in the point of maximum increase with the growth energy $\sqrt{s}$.

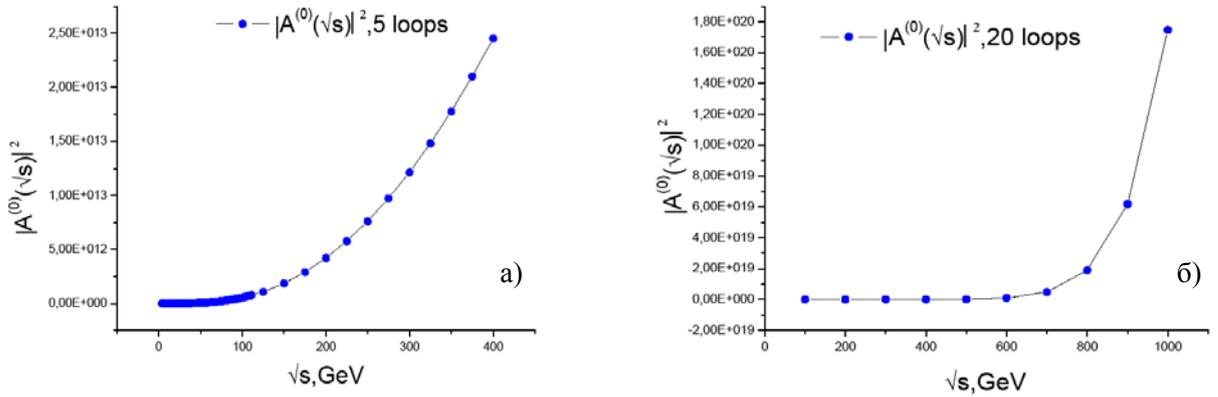

Fig.12. The scattering amplitude squared modulus, that correspond to the diagram Fig. 5 in the maximum $\left|A^{(0)}\left(\sqrt{s}\right)\right|^2$, as a function of energy $\sqrt{s}$ of secondary particles Fig. 5 : a) for $n=5$, b) for $n=20$.

Therefore, the same mechanism of cross-section growth, which has been discovered in the simple scalar models, also takes place in the model with gluons loops. The same concerns the peculiar feature of rapidities in the point of maximum, i.e. they produce an arithmetic progression, as can be clearly seen from Fig. 13.

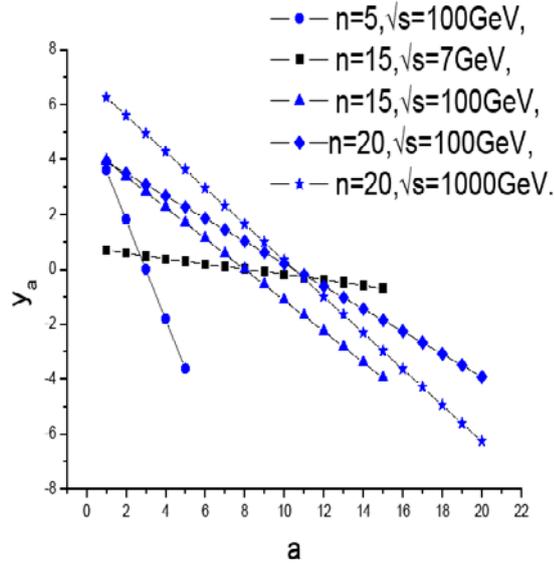

Fig.13. Typical plot of the rapidity as a function of vertex number in the point of maximum.

The value of scalar square of four momentum $K^2$ is given in Figure 14 as a function of loop number for different energies $\sqrt{s}$.

In Figure 14 we can see graphs of dependencies for the different values $K^2$ of loops (see Fig.4), calculated using rapidities at the point of maximum with different values $\sqrt{s}$

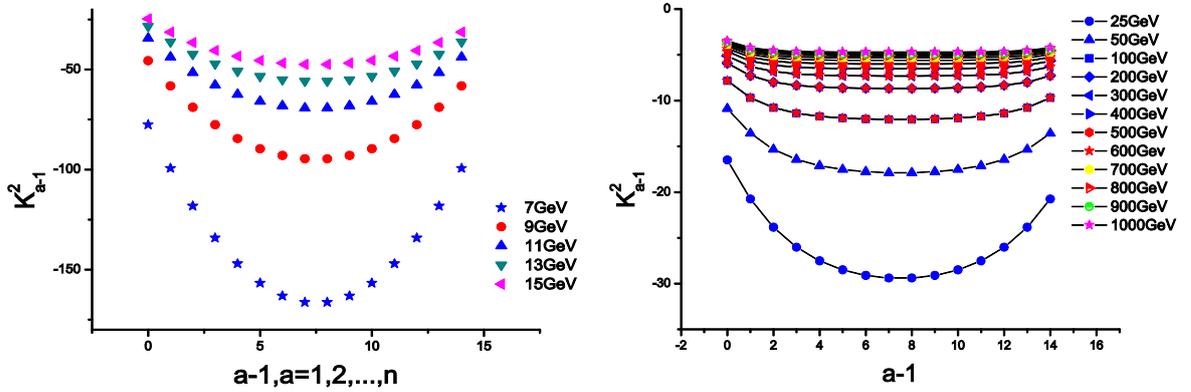

Fig.14 The scalar square of four-momenta $K^2_{a-1}$ entering the loop in Fig. 4 (herewith, the loop containing the a-th vertex is labeled as "a-1") as a function of number of loop at different energies $\sqrt{s}$.

These graphs are quite similar to the ones in Figure 6 of Ref. (2), obtained for the simplest multi-peripheral "comb" diagrams in $\varphi^3$ theory. The origin of the analogy between the model, based on the diagrams with loops in QCD, and the one based on loop-less diagrams in $\varphi^3$ theory can be explained from the following concerns.

At the numerical maximization it has been observed that the outcome only slightly differs from the one, which would have been obtained if the multiplier corresponding to each loop (16) would be replaced with the expression for the inverse Jacobian only (12). This expression enters (16) as a factor in front of the integral in the real part, and before the logarithm of the imaginary part. Above illustrated graphs are shown in Figure 15.

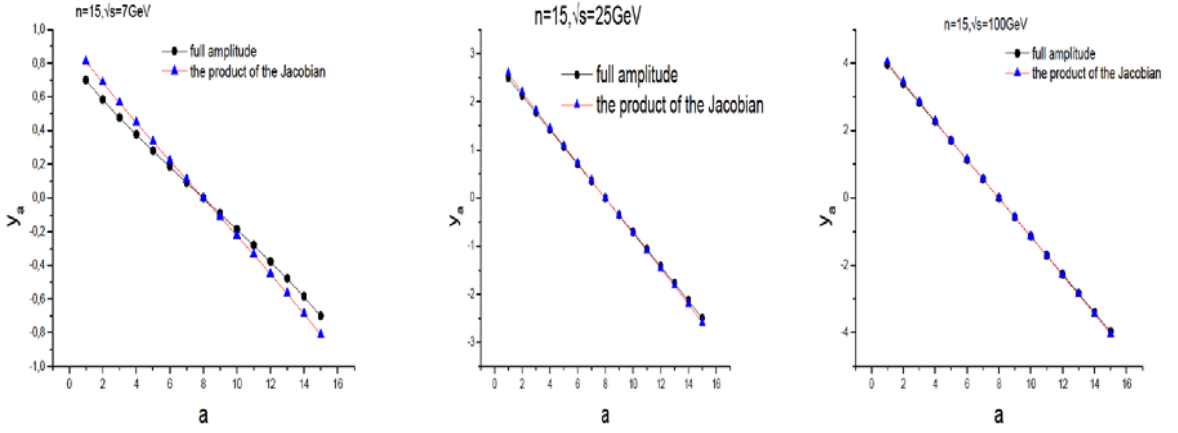

Figure 15 Comparison of the results of maximization the total amplitudes (circles) and the products of Jacobians only, given at different energies.

Thus, instead of the full scattering amplitude, with good accuracy one can maximize the logarithm of the product of the Jacobians. Dropping out the multipliers and the constant summands, which are irrelevant to the maximization, this logarithm can be reduced to the form:

$$\tilde{A} = -\sum_{a=1}^{N} \ln\left(\left(K_{a-1} p_a\right)^2 - K_{a-1}^2 p_a^2\right) \qquad (23)$$

As shown in Figure 14 in the vicinity of the point of maximum the values $K_{a-1}^2$ are almost equal between themselves.
Namely

$$K_{a-1}^2 \approx \left(K_{a-1} - p_a\right)^2. \qquad (24)$$

Consequently

$$\left(K_{a-1} p_a\right) \approx \frac{p_a^2}{2} = 2, \qquad (25)$$

Here the mass-shell condition is taken into account, and momenta are nondimensionalized with mass of secondary particle.

These concerns are confirmed by the results of the direct calculation of the values $\left(K_{a-1} p_a\right)$, given in Fig.16

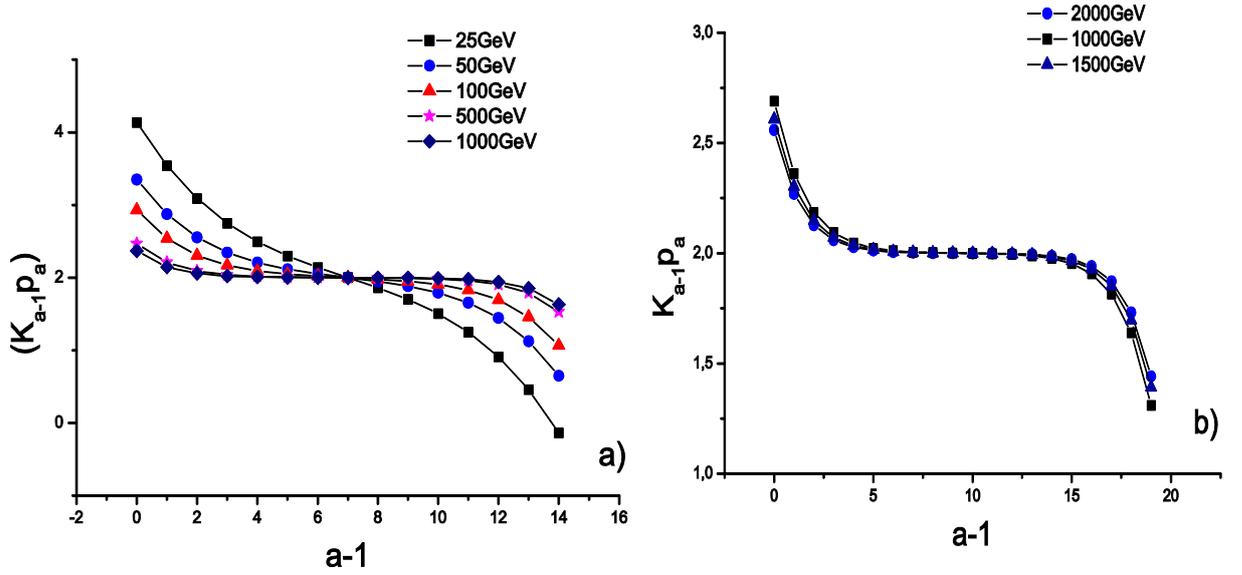

Figure 16. The value of $(K_{a-1}p_a)$ as a function of loop number at the point of maximum, given at different energies $\sqrt{s}$: a) $n = 15$, b) $n = 20$.

Given the approximation (25) instead of (23) we get:

$$\tilde{A} = -N\ln(4) - \sum_{a=1}^{N}\ln\left(1 - K_{a-1}^2\right). \qquad (26)$$

This expression, up to the first term, which is nonessential for the maximization, coincides with the expression that occurs in the simplest version of $\varphi^3$ theory for the "comb" diagrams, and which was previously employed in Refs. [1-5].

## 4. Conclusions and discussion

The main conclusion from the above results is that the inelastic process of exchange with two massless gluons in QCD (see Fig. 6) is equivalent to exchange with one massive particle in a scalar $\varphi^3$ theory. However, we by no means employ the features of QCD, i.e. such an outcome will occur in any arbitrary non-Abelian gauge theory, which features the massive secondary particles.

Moreover, as one can see from the results in Figure 14, within the considered model, the mechanism of cross-sections growth, which was discovered earlier for the effective scalar theories [2, 5], remains. The essence of this mechanism is as follows.

The scalar square of each four-momentum $K_{a-1}^2$ is negative. This means that the three-dimensional momentum, which runs through the loop, can't be reduced to zero in any reference frame. The value of $\left|K_{a-1}^2\right|$ is equal to the square of this three-dimensional momentum in the reference frame, in which the momentum is minimal. According to the uncertainty relation, the smaller is this momentum the bigger is the spatial region, in which one can find during the measurement the particles involved in the process, described by loop (see Figure 6), or the particle that transfers the four-momentum $K_{a-1}$ along the scalar "comb". The growth of this region, which is a consequence of results shown in Figure 14 defines the growth of cross-section of primary particles scattering.